\documentclass[epsf,usegraphicx]{mn2e}

\usepackage{xspace}

\title[The X-ray spectrum of RX J1914.4+2456 revisited]
{The X-ray spectrum of RX J1914.4+2456 revisited}

\author[]
{Gavin Ramsay\\
Armagh Observatory, College Hill, Armagh, BT61 9DG, Northern Ireland\\ 
}

\date{Accepted 2007 November 15.  Received 2007 November 14; in
original form 2007 October 9}

\begin{document}
\outer\def\gtae {$\buildrel {\lower3pt\hbox{$>$}} \over 
{\lower2pt\hbox{$\sim$}} $}
\outer\def\ltae {$\buildrel {\lower3pt\hbox{$<$}} \over 
{\lower2pt\hbox{$\sim$}} $}
\newcommand{\ergscm} {ergs s$^{-1}$ cm$^{-2}$}
\newcommand{\ergss} {ergs s$^{-1}$}
\newcommand{\ergsd} {ergs s$^{-1}$ $d^{2}_{100}$}
\newcommand{\pcmsq} {cm$^{-2}$}
\newcommand{\ros} {\sl ROSAT}
\newcommand{\chan} {\sl Chandra}
\newcommand{\xmm} {\sl XMM-Newton}
\def\rchi{{${\chi}_{\nu}^{2}$}}
\newcommand{\Msun} {$M_{\odot}$}
\newcommand{\Mwd} {$M_{wd}$}
\def\Mdot{\hbox{$\dot M$}}
\def\mdot{\hbox{$\dot m$}}
\newcommand{\teff}{\ensuremath{T_{\mathrm{eff}}}\xspace}

\maketitle

\begin{abstract}
It has been proposed that RX J1914.4+2456 is a stellar binary system
with an orbital period of 9.5 mins. As such it shares many similar
properties with RX J0806.3+1527 (5.4 mins). However, while the X-ray
spectrum of RX J0806.3+1527 can be modelled using a simple absorbed
blackbody, the X-ray spectrum of RX J1914.4+2456 has proved difficult
to fit using a physically plausible model. In this paper we re-examine
the available X-ray spectra of RX J1914.4+2456 taken using {\xmm}. We
find that the X-ray spectra can be fitted using a simple blackbody and
an absorption component which has a significant enhancement of neon
compared to the solar value.  We propose that the material in the
inter-binary system is significantly enhanced with neon. This makes
its intrinsic X-ray spectrum virtually identical to RX
J0806.3+1527. We re-access the X-ray luminosity of RX J1914.4+2456 and
the implications of these results.

\end{abstract}


\section{Introduction}

The X-ray source RX J1914.4+2456 (hereafter RX J1914+24) has been the
subject of much debate since its discovery during the {\ros} all-sky
survey. A number of competing models have been put forward to account
for its observed properties, but all of them involve a stellar binary
system. The models can be split into accretion and non-accretion
models. The non-accretion model is the unipolar inductor (UI) model
where dissipation of large electrical currents heat the magnetic white
dwarf (Wu et al 2002).  It shares many similar observational
characteristics to the X-ray source RX J0806.3+1527 (hereafter RX
J0806+15, see Cropper et al 2004a for a review).

One of the main uncertainties to understanding the nature of RX
J1914+24 is accurately determining its X-ray luminosity,
$L_\mathrm{X}$. In the UI model, $L_\mathrm{X}$ is proportional to the
rate of change of the orbital period and the degree of
asynchronisation between the binary orbital period and the primary
star (Wu et al 2002, Dall'Osso et al 2007).

\begin{table*}
\begin{center}
\begin{tabular}{lrrrrrrrrr}
\hline
Revolution & Date of & \multicolumn{2}{c}{EPIC PN} & 
\multicolumn{2}{c}{EPIC MOS1} & 
\multicolumn{2}{c}{EPIC MOS2} & RGS & Flux\\
& Observation & Mode &  Duration & Mode & Duration & Mode & Duration & 
Duration & \ergscm\\ 
\hline
0718 & 2003-11-09 & sw & 6641 & timing & 8933 & sw & 8993 &  9880 & 
$1.25\times10^{-12}$ \\
0721 & 2003-11-15 & sw & 2888 & timing & 3865 & sw & 3865 & 4000  &
$1.20\times10^{-12}$ \\
0880 & 2004-09-28 & ff & 11869 & ff & 14954 & ff & 14954 &  16954 &
$1.27\times10^{-12}$ \\
0882 & 2004-10-02 & ff & 15274 & ff & 18419 & ff & 18419 & 18880  &
$1.40\times10^{-12}$ \\
\hline
\end{tabular}
\end{center}
\caption{The log for the observations of RX J1914+24 made using {\xmm}.
We show the mode the detector was in where `sw' refers to `small window', 
`ff' to `full frame' and the duration of `good' time -- ie excluding 
time intervals of enhanced background. The RGS detectors were configured 
in `Spectroscopy' mode. In the last column we show the observed integrated
flux in the 0.2--10keV energy band using the EPIC pn detector data 
and fitting an absorbed blackbody plus broad emission line spectral model.}
\label{log}
\end{table*}

In practise it has been difficult to get an accurate value for
$L_\mathrm{X}$. This is partly due to the fact that RX J1914+24 is
highly reddened. The other factor is that its X-ray spectrum is rather
unusual and therefore difficult to determine the underlying emission
model. X-ray spectra obtained using {\xmm} are not well fitted using a
simple absorbed blackbody, showing large residuals near 0.7 keV
(Ramsay et al 2005). Ramsay et al (2005) found that an absorbed
blackbody with a broad emission line centered at 0.59keV gave a much
improved goodness of fit and for a distance of 1kpc implied
$L_\mathrm{X}\sim10^{35}$ \ergss.

Using two further longer series of observations of RX J1914+24 also
taken using {\xmm}, Ramsay et al (2006) found that an absorbed
low-temperature thermal plasma model with an edge at 0.83keV gave a
significantly improved fit compared to the previous best model fit.
For a distance of 1kpc this optically thin emission model gave
$L_\mathrm{X}\sim10^{33}$ \ergss. On the other hand if the distance
was much lower than 1 kpc (Steeghs et al 2006, Barros et al 2007) then
$L_\mathrm{X}$ could be as low as $\sim3\times10^{31}$ \ergss --
giving a range in $L_\mathrm{X}$ of 4 orders of magnitude!

RX J1914+24 has been observed using {\xmm} at 4 separate epochs (Table
\ref{log}). An analysis of the data taken using the EPIC detectors
have been presented in Ramsay et al (2005) (from the first two epochs)
and in Ramsay et al (2006) (from the last two epochs). In this paper
we examine the data obtained using the RGS detectors; re-examine the
data obtained using the EPIC detectors using more recent calibration
data, and also re-examine the models used to fit the data.

\section{Observations}

The data were processed using {\xmm} SAS v7.0 (the data presented
previously were processed using v6.0 and v6.5 in Ramsay et al 2005 and
Ramsay et al 2006 respectively). In our analysis we excluded time
intervals of high particle/solar background (a significant issue in
the second epoch observation).

For those observations when the EPIC data were in full frame mode, we
excluded events from the central core of the psf (using an aperture of
10$^{''}$ in radius) in order that pile-up was not significant. We did
not extract spectra from the timing mode data since the spectral
calibration is not as well defined as for the other modes. For the RGS
data, we extracted spectra which included the source and background,
and a background spectrum separately. We grouped the EPIC spectra so
that each bin had a minimum of 40 counts. Since the RGS spectra from
the individual epochs were relatively low, we co-added the spectra
from the RGS1 and RGS2 detectors obtained using the 3rd and 4th epoch
observations (the first two epochs had much shorter exposures). We
used the {\tt SAS} task {\tt rgscombine} and then grouped each
spectrum so that each bin of each spectrum had a minimum of 20 counts
per bin.

To determine the observed X-ray flux at each epoch, we used data taken
using the EPIC pn detector. We fitted the integrated X-ray spectra
using an absorbed blackbody plus broad emission line. We show the
integrated observed flux in the 0.2--10keV energy band using in Table
\ref{log} (the observed flux is only weakly sensitive to the model
used). This shows that the observed flux varied by 17\% between the 4
observation epochs.

The X-ray data folded on the 569 sec period shows a distinctive
`on--off' behaviour, with the X-ray flux being off for approximately
half the 569 sec period (Cropper et al 1998). There is a sharp rise in
flux which is followed by a slower decline from maximum
brightness. Ramsay et al (2005) showed evidence using the two shorter
duration observations that the spectrum gets softer during this
decline phase. Using the 3rd and 4th longer series of observations we
confirm this finding. Therefore we have obtained a spectrum which
covers the `bright phase' which we define to be $\phi$=0.08--0.38
where $\phi$=0.0 is defined as the start of the sharp rise in X-ray
flux.

\section{The RGS spectra}

\begin{table}
\begin{center}
\begin{tabular}{lr}
\hline
Model & \rchi \\
      & (dof) \\
\hline
tbabsvmekaledge & 2.18 (93) \\
tbabsbb & 2.12 (95) \\
tbvarabsbb & 1.28 (93)  \\
tbabsbbgau (abs)  & 1.20 (92) \\
tbabsbbgau (emi)  & 1.14 (91) \\
\hline
\end{tabular}
\end{center}
\caption{The fits to the bright phase RGS spectrum taken from data in {\xmm}
orbits 0880 and 0882. 
The models noted in the first column refer
to the models in {\tt XSPEC}: tbabs -- T\"{u}bingen Boulder absorption 
ISM model (Wilms et al 2000); 
tbvarabs -- T\"{u}bingen Boulder absorption ISM model with 
variable abundances; bb -- blackbody;
gau -- a Gaussian component in emission (emi) and absorption (abs); vmekal -- 
a thermal plasma model with non-solar abundances; edge - an absorption edge;
In the second column we show the \rchi \hspace{1mm} and (degrees of freedom).}
\label{rgsfits}
\end{table}

We extracted RGS spectra from the bright phase using the 3rd and 4th
epoch observations. In fitting the spectra, we used a blackbody, a
blackbody plus a Gaussian component both in absorption and emission
and a multi-temperature thermal plasma plus edge model. In the work of
Ramsay et al (2005, 2006), the absorption model which was used was the
`{\tt wa}' neutral absorption model found in the {\tt XSPEC} fitting
package (Arnaud 1996). Here, we use the T\"{u}bingen--Boulder
absorption ISM model and abundances (Wilms, Allen \& McCray 2000)
which incorporates advances in atomic cross-sections and other
physical parameters compared to the {\tt wa} model (Morrison \& 
McCammon 1983).  We used this model implemented into {\tt XSPEC} as the
{\tt tbabs} model (which assumes an interstellar medium of solar
abundance) and the {\tt tbvarabs} model (which allows the abundance of
each element to vary).

We show the goodness of fit to the RGS spectrum using the different
models in Table \ref{rgsfits}. As was found by Ramsay et al (2005) a
simple absorbed blackbody model gives a very poor fit. Ramsay et al
(2006) found that a low temperature thermal plasma plus edge model
gave a good fit to the spectrum obtained using the EPIC pn
detector. Using the RGS data we can rule this model out. The
temperature of the plasma determined using the EPIC pn detector is
very low, ($<1$keV), which would result in strong X-ray emission lines
-- these lines are not detected in the RGS data.

\begin{figure}
\begin{center}
\setlength{\unitlength}{1cm}
\begin{picture}(14,6)
\put(-0.5,-0.5){\includegraphics{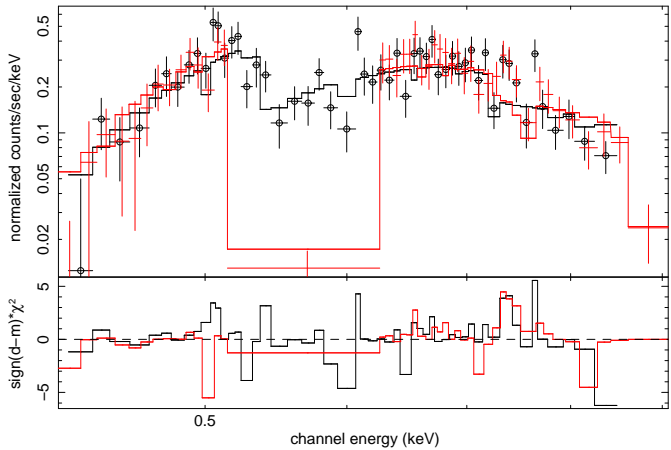}}
\end{picture}
\end{center}
\caption{The fits to the RGS spectra bright phase spectra. The spectra
from each detector (RGS1 - datapoints plotted as black circles, RGS2
red points) taken in the 3rd and 4th epoch observations.  The solid
lines show the best fit using an absorbed blackbody model (the darker
line through the RGS1 data and lighter, or red, line for the RGS2
deta), where the absorption component has variable abundances.}
\label{spectra}
\end{figure}

This leaves three models -- a blackbody with either an absorption or
emission component, or a blackbody with an absorption component which
has abundance different to solar composition. We show the RGS spectrum
together with the best fit using an absorbed blackbody, where the
absorption component has variable abundances, in Figure \ref{spectra}.

\section{The EPIC pn bright phase spectra}

We extracted bright phase spectra from each epoch. We fitted models
consisting of a blackbody and a Gaussian line in both emission and
absorption, and also a blackbody with an absorption component with
variable abundances. We show the goodness of fits to these spectra
using these three models in the top panel of Table
\ref{epic_fits}. (For the second epoch observation, which had a short
good exposure we fixed the model parameters at the best fit parameters
determined in fourth observation apart from the absorption column
density and the normalisation parameters). We find that the models
which include a Gaussian in absorption or emission give formally good
fits (95 \% confidence) while the model with an interstellar component
with variable interstellar abundance gives a \rchi $\le1.0$.

We show the observed flux in the 0.2-10keV energy band and also the
unabsorbed flux in the 0.01--10keV energy band in the bottom panel of
Table \ref{epic_fits}. The observed flux at each epoch as derived from
the various models is consistent to within 7\%. In contrast, the
derived unabsorbed flux varies by more than an order of magnitude.

\begin{table*}
\begin{center}
\begin{tabular}{lrrrr}
\hline
Model & 0718 & 0721 & 0880 & 0882 \\
      & \rchi (dof)  & \rchi (dof) & \rchi (dof) &  \rchi (dof)\\
\hline
tbabsbbgau (emi)  & 0.90 (51) & 0.90 (27) & 1.24 (70) & 1.02 (88) \\
tbabsbbgau (abs)  & 0.93 (51) & 1.43 (23) & 0.83 (70) & 1.07 (88) \\
tbvarabsbb & 0.82 (53)  & 0.92 (27) & 1.00 (72) & 0.88 (88) \\
\hline
\end{tabular}
\begin{tabular}{lrrrrrrrr}
\hline
Model & \multicolumn{2}{c}{0718} & \multicolumn{2}{c}{0721} &
 \multicolumn{2}{c}{0880} & \multicolumn{2}{c}{0882}\\
      & Flux$^{o}$ & Flux$^{u}$ &  Flux$^{o}$ & Flux$^{u}$ &
       Flux$^{o}$ & Flux$^{u}$ &  Flux$^{o}$ & Flux$^{u}$ \\
\hline
tbabsbbgau (emi)  & 8.24$\times10^{-12}$ & 7.35$\times10^{-10}$ 
                  & 8.75$\times10^{-12}$ & 8.80$\times10^{-10}$ 
                  & 3.45$\times10^{-12}$ & 2.60$\times10^{-10}$ 
                  & 3.62$\times10^{-12}$ & 6.19$\times10^{-10}$ \\
tbabsbbgau (abs)  & 8.54$\times10^{-12}$ & 1.82$\times10^{-10}$ 
                  & 8.97$\times10^{-12}$ & 2.20$\times10^{-10}$ 
                  & 3.50$\times10^{-12}$ & 2.68$\times10^{-10}$ 
                  & 3.63$\times10^{-12}$ & 2.42$\times10^{-10}$ \\
tbvarabsbb        & 8.27$\times10^{-12}$ & 2.90$\times10^{-9}$ 
                  & 8.73$\times10^{-12}$ & 1.80$\times10^{-9}$ 
                  & 3.49$\times10^{-12}$ & 3.38$\times10^{-10}$ 
                  & 3.62$\times10^{-12}$ & 1.34$\times10^{-9}$ \\
\hline
\end{tabular}
\end{center}
\caption{Top panel: The fits to the bright phase EPIC pn spectrum
obtained using {\xmm} at four epochs. The models noted in the first
column are defined in the caption for Table 2.  In the following
columns we show the goodness of fit (\rchi and degrees of
freedom). Bottom panel: we show the observed flux, Flux$^{o}$, in the
0.2--10keV energy band, and the implied unabsorbed flux, Flux$^{u}$,
in the 0.01--10keV energy band for all four epochs and the three
spectral models.}
\label{epic_fits}
\end{table*}

\begin{table*}
\begin{center}
\begin{tabular}{lrrrrr}
\hline
{\sl XMM} & $N_\mathrm{H}$  & $kT_{bb}$ & E & $\sigma$ & EW\\
Orbit  & (10$^{21}$ \pcmsq)   & (eV)    & (keV) & (keV) & (eV) \\ 
\hline
\multicolumn{6}{c}{Gaussian component in emission}\\
\hline
0718 & 4.2$^{+0.5}_{-0.7}$ & 64.9$^{+1.9}_{-4.6}$ & 0.64$^{+0.02}_{-0.03}$ &
0.075$^{+0.018}_{-0.014}$ & 182$^{+104}_{-55}$\\
0880 & 5.7$^{+0.1}_{-0.2}$ & 57.5$^{+1.8}_{-2.0}$ & 0.63$\pm0.01$ & 
0.093$^{+0.004}_{-0.005}$ & 152$^{+15}_{-20}$\\
0882 & 5.1$^{+0.1}_{-0.7}$ & 63.9$^{+1.1}_{-4.7}$ & 0.68$\pm0.01$ & 
0.06$^{+0.02}_{-0.01}$ & 147$^{+61}_{-26}$\\
\hline
\multicolumn{6}{c}{Gaussian component in absorption}\\
\hline
0718 & 2.9$^{+0.3}_{-0.2}$ & 110$^{+18}_{-10}$ & 0.86$^{+0.02}_{-0.47}$ &
0.111$^{+0.006}_{-0.006}$ & -216$^{+53}_{-247}$\\
0880 & 4.9$^{+0.7}_{-0.7}$ & 80.7$^{+17.7}_{-10.2}$ & 0.87$^{+0.04}_{-0.10}$ & 
0.12$^{+0.03}_{-0.03}$ & -148$^{+64}_{-226}$\\
0882 & 6.2$^{+0.5}_{-0.9}$ & 70.6$^{+6.2}_{-2.2}$ & 0.84$^{+0.02}_{-0.09}$ & 
0.105$^{+0.03}_{-0.02}$ & -104$^{+39}_{-134}$\\
\hline
\end{tabular}
\end{center}
\caption{The spectral parameters for fits to the bright phase spectra using
an absorbed blackbody model where we have allowed the abundance of the 
absorption component to vary from solar. The duration of the EPIC pn spectrum
from {\xmm} orbit 0721 was relatively short and hence the spectral parameters 
are not strongly constrained.}
\label{fits}
\end{table*}

We show the spectral parameters derived for the blackbody plus a
Gaussian component in Table 4.  The spectral parameters derived using
a blackbody plus Gaussian line in emission gives consistent results,
with a blackbody temperature $kT\sim$62 eV and
$N_\mathrm{H}\sim5\times10^{21}$ \pcmsq. The center of the Gaussian component
is $\sim$0.65 keV and has a width $\sim$80 eV and an equivalent width
$\sim$160 eV. In contrast, the spectral parameters as derived using
the model with the Gaussian line in absorption, show a greater
spread. However, the temperature of the blackbody component is hotter
than in the previous model. The center of the absorption line is
$\sim$0.86 keV and has a width $\sim$0.11 keV.

Since the variable abundance absorption model was not used by Ramsay
et al (2005, 2006) we now go onto investigate this model in more
detail.  We fit the spectra using an absorbed blackbody model where we
allowed the abundance of the absorbing material to vary from solar. We
did this by allowing each element to vary from solar one at a time. If
there was no significant change in the goodness of fit, we re-set the
abundance for that element back to solar and fixed this parameter. We
did this for the EPIC pn and RGS spectra from {\xmm} orbits 0880 and
0882 individually.  We show the spectral parameters for these fits in
Table \ref{tbvarabs_fits}.  We find that their X-ray spectra can be
well fitted using blackbody with an interstellar absorption model
which has significantly increased amounts of Neon. There is some
evidence that Iron could have non-solar abundances.

We also fitted the EPIC pn spectra taken from {\xmm} orbits 0718 and
0721 using this same model. As mentioned before, for the data taken in
the orbit 0721 observations, we fixed the spectral parameters at their
best fit parameters determined in orbit 0882 (apart from the
absorption column density and the blackbody normalisation). We show the
spectral parameters derived from these spectra in Table
\ref{tbvarabs_fits}. We find these spectra are also consistent with
the abundance of neon being significantly enhanced in these epochs as
well.

\begin{table*}
\begin{center}
\begin{tabular}{lrrrrrrrrr}
\hline
{\sl XMM} & Detector & $N_\mathrm{H}$  & $kT_{bb}$ & N & Ne & Fe & \rchi\\
Orbit  &          & (\pcmsq)   & (eV)    & ($Z_{\odot}$) & ($Z_{\odot}$) & (dof) \\
\hline
0718 & pn & 6.1$^{+0.2}_{-0.2}$ & 65$^{+13}_{-9}$ & & 9.5$_{-1.5}^{+1.9}$ & &0.90 (50)\\ 
\hline
0721 & pn & 5.6$^{+0.3}_{-0.3}$ & 65 & 0.02 & 7.8 & 0.0 & 0.84 (22)\\ 
\hline
0880 & pn & 4.3$^{+1.4}_{-1.0}$ & 72$^{+12}_{-8}$ & & 20$_{-7}^{+12}$ & & 
1.00 (72)\\
0880 & rgs & 3.6$^{+1.0}_{-0.6}$ & 67$^{+7}_{-5}$ & & 39$_{-4}$ & 
2.6$^{+1.0}_{-0.9}$ &  1.09 (162)\\
\hline
0882 & pn & 6.8$^{+0.2}_{-0.1}$ & 65$^{+9}_{-5}$ & $<0.43$& 7.8$^{+1.1}_{-1.0}$ & 
$<$0.26 & 0.88 (88)\\
0882 & rgs & 5.2$^{+1.4}_{-1.1}$ & 66$^{+4}_{-4}$ & & 20$^{+3}_{-5}$ & &
1.03 (192)\\
\hline
\end{tabular}
\end{center}
\caption{The spectral parameters for fits to the bright phase spectra
using an absorbed blackbody model where we have allowed the abundance
of the absorption component to vary from solar. Since the duration of
the spectrum taken in orbit 0721 was short, the spectral parameters
were fixed at their values determined in orbit 0882 apart from the
total absorption column density and normalisation of the blackbody
component.}
\label{tbvarabs_fits}
\end{table*}

\section{An evaluation of the spectral models}

We know go onto to discuss the physical plausibility of the three
spectral models which we have used. Namely, the absorbed blackbody
with an absorption component and an emission component, and the
blackbody with absorption component with non-solar abundances.

\subsection{A blackbody with Gaussian absorption component}
\label{ins}

Isolated neutron stars (INS) with high magnetic fields have been found
to show absorption lines which have been attributed to either a proton
cyclotron line or an electron cyclotron line, see Zane et al (2005),
Haberl (2007) and Schwope et al (2007).  The similarity between the
spectral parameters of RX J1914+24 and some INS is quite striking. For
instance, the width of the line and its equivalent width as measured
in RX J1914+24 is very similar to that of RBS 1223 (Schwope et al
2007). Also the variation in the observed flux in the energy range
0.35--1.5keV of the INS RX J0720.4--3125 is 20\% (cf 17 \% between the
{\xmm} observations of RX J1914+24) showing that non-accreting sources
can show a significant variation in their X-ray flux (the data on RX
J0720.4--3125 were extracted from the {\xmm} archive).

There are, however, some very significant differences between the
observed properties of RX J1914+24 and that of INS. The first is that
their spin-periods are in the range $\sim3-12$ sec - these are very
much shorter than the period seen in RX J1914+24 (569 sec). The second
is the luminosity difference - INS show X-ray luminosities \ltae
$10^{31}$ \ergss. Using the inferred unabsorbed fluxes as derived
using the blackbody plus Gaussian absorption line, we find that RX
J1914+24 would have to be at a distance of 20 pc to give a comparable
luminosity.  If RX J1914+24 was so close, we would expect to detect a
significant proper motion which has not been observed (Israel et al
2002).  The third difference is the change in the period. The 569 sec
period in RX J1914+24 has been found to be decreasing, while the
period of the two INS which have been found to show a change in their
period are increasing (Cropper et al 2004b, Kaplan \& van Kerkwijk
2005a, Kaplan \& van Kerkwijk 2005b). A fourth difference is the
optical brightness, with INS being typically $B\sim$26 (see the
compilation in Haberl 2007), while RX J1914+24 is $B\sim$21.  We
conclude RX J1914+24 is not an isolated neutron star. At this point,
whilst we cannot rule out the presence of absorption features in the
X-ray spectrum of RX J1914+24, we do not consider it the most likely
model to explain its X-ray spectrum.

\subsection{A blackbody with Gaussian emission component}

A blackbody with an additional broad emission line does, on first
sight, seem rather contrived. However, such a feature has been claimed
to be present in the relatively low resolution {\sl ASCA} spectra of a
number of X-ray ultra-compact binaries with neutron star primaries (eg
Juett, Psaltis \& Chakrabarty 2001). The line center of the emission
line is remarkably similar in RX J1914+24 and the four sources shown
described in Juett et al (2001).  One notable difference is that the
temperature of the blackbody component in the X-ray UCBs is much
hotter -- several hundred eV as opposed to $\sim$60 eV -- and the fact
that they are hard X-ray sources being detected at energies up to many
10's of keV. This is the result of the primary being a neutron star as
opposed to a white dwarf.

It was claimed that this broad emission feature was due to the
superposition of a number of unresolved emission lines. However, when
one of the sources observed using {\sl ASCA} was observed using the
{\sl Chandra} Low Energy Grating Spectrometer, it failed to detect any
emission features which could give rise to a broad emission feature in
low resolution spectra. We now address one possible reason for this.

\subsection{A blackbody with variable abundances in the absorption model}

Juett et al (2001) found that good fits to {\sl Chandra} spectra of
neutron star UCBs were obtained if the absorption model had non-solar
abundances. In particular, they found a high relative abundance of
neon and suggested that this over-abundance was located in the
inter-binary system.  However, further work (eg Juett \& Chakrabarty
2005) showed that for individual sources, the Ne/O ratio showed
evidence for variability from epoch to epoch, which they attributed to
source variability. This implied that the abundances could not be used
to determine the composition of the mass donating star.

There is a clear similarity between the neutron star UCBs described by
Juett et al and RX J1914+24. In each observation of RX J1914+24, there
is clear evidence that the absorption component has an overabundance
of neon.

\section{Discussion}

For the reasons outlined in \S \ref{ins}, we rule out RX J1914+24
being an isolated neutron star. Since all the known neutron star UCBs
have X-ray emission extending up to many 10's of keV we also rule out
an accreting neutron star UCB model. We cannot rule out that a neutron
star is in a binary system where a secondary star was not filling its
Roche Lobe. In this scenario, an X-ray bright system would have to be
powered by UI.

It is highly unlikely that the line of sight absorption to RX J1914+24
has a chance enhancement of neon. It is much more likely that this
overabundance is concentrated in the binary system. Juett \&
Chakrabarty (2005) noted that for some neutron star UCBs the Ne/O
abundance varied from epoch to epoch and hence the observations could
not be used to determine the abundance of the secondary, mass-donating
star, in the binary system. In the case of RX J1914+24 there is clear
evidence for a significant over-abundance of neon in the absorption
component at each epoch. At this stage it is not clear if this
over-abundance is due to circumbinary material left over from a
previous stage in the binary formation process or can give us a direct
insight into the chemical composition of the secondary star (if
accretion is occurring).

What are the implications of our findings regarding the X-ray
luminosity of RX J1914+24? Steeghs et al (2006) discuss the extinction
and distance estimates to RX J1914+24. While the distance is rather
uncertain, it is likely that it is greater than $\sim$1 kpc.  We can
rule out the lower estimates ($L_\mathrm{X}\sim10^{33}$ \ergss for a distance
of 1 kpc) which were derived using a low temperature thermal plasma
model.  Taking the unabsorbed bolometric fluxes derived using the
blackbody with absorption component with variable abundances and
assuming a distance of 1 kpc we find $L_\mathrm{X}=2\times10^{34} -
1.6\times10^{35}$ \ergss.

Dall'Osso et al (2007) made a detailed investigation of the UI model
in the context of RX J1914+24 and RX J0806+15. They predicted that for
low luminosities, $L_\mathrm{X}\sim10^{33}$ \ergss, the asynchronism between
the orbit and the magnetic star in RX J1914+24 would have to be
$\alpha\sim0.9-$0.98, where $\alpha=\omega_{1}/\omega_{o}$, and
$\omega_{1}$ is the rotation frequency of the primary star and
$\omega_{o}$ is the orbital frequency. Unless RX J1914+24 was located
at a distance significantly less than 1kpc, we can rule these low
values of asynchronism. For high luminosities ($L_\mathrm{X}=10^{34-35}$
\ergss), Dall'Osso et al (2007) predicted that an asynchronism of a
few was required ($\alpha\sim$ 4). For an observed period of 569 sec,
$\alpha=2-10$ gives a predicted period of $\sim$60-300 sec. There is
no evidence for power at these periods in the power spectra of the
X-ray light curves (Ramsay et al 2006).

\section{Summary}
 
Until now the nature of the emission source that powers the X-ray
spectrum of RX J1914+24 has been far from clear. In this paper we have
shown that it can be well modelled using a simple blackbody model with
an absorption component which has non-solar abundances, in particular,
an enhancement of neon.

Since the X-ray light curves of RX J1914+24 and RX J0806+15 are
practically identical, it suggests that their X-ray emission source is
the same. The fact that their X-ray spectra were apparently different
(albeit both being soft) was therefore perplexing. Our result showing
that the emission source is the same for both RX J1914+24 and RX
J0806+15 is therefore very attractive.  Indeed their temperatures are
virtually identical -- we obtain a mean value of $kT\sim67$ eV for RX
J1914+24 compared to $kT\sim65$ eV for RX J0806+15 (Israel et al
2003).

The difference between the X-ray spectrum of RX J1914+24 and RX
J0806+15 is that the absorption component of RX J1914+24 has enhanced
neon abundance. A further investigation of the optical spectrum of RX
J1914+24 to search for neon features is strongly encouraged.

\section{Acknowledgements}

I will wish to thank Mark Cropper, Pasi Hakala and Peter Wheatley for
useful discussions. Armagh Observatory is grant aided by the
N. Ireland Dept. of Culture, Arts and Leisure.


\begin{thebibliography}{}

\bibitem{}Arnaud, K. A., 1996, Astronomical Data Analysis Software and 
Systems V, eds Jacoby, G., Barnes, J., p17, ASP Conf Series, 101
\bibitem{}Barros, S. C. C., Marsh, T. R., Dhillon, V. S., Groot, P. J., 
Littlefair, S., Nelemans, G., Roelofs, G., Steeghs, D., Wheatley, P. J., 2007,
MNRAS, 374, 1334
\bibitem{}Cropper, M., Harrop-Allin, M. K., Mason, K. O., Mittaz, J. P. D., 
Potter, S. B., Ramsay, G., 1998, MNRAS, 293, L57
\bibitem{}Cropper, M., Ramsay, G., Wu, K., Hakala, P., 2004a, In Proc `Magnetic 
Cataclysmic Variables', IAU Colloquium 190, ASP Conference Proceedings, Vol. 
315., p324
\bibitem{}Cropper, M., Haberl, F., Zane, S., Zavlin, V. E., 2004b, MNRAS,
351, 1099
\bibitem{}Dall'Osso, S., Israel, G. L., Stella, L., 2007, A\&A, 464, 417
\bibitem{}Haberl, F., 2007, Ap\&SS, 308, 181
\bibitem{}Juett, A. M., Psaltis, D., Chakrabarty, D., 2001, ApJ, 560, L59
\bibitem{}Juett, A. M., Chakrabarty, D., 2005, ApJ, 627, 926
\bibitem{}Israel, G. L., 2002, A\&A, 386, L13
\bibitem{}Israel, G. L., 2003, ApJ, 598, 492
\bibitem{}Kaplan, D. L., van Kerkwijk, M. H., 2005a, ApJ, 628, L45
\bibitem{}Kaplan, D. L., van Kerkwijk, M. H., 2005b, ApJ, 635, L65 
\bibitem{}Morrison, R., McCammon, D., 1983, ApJ, 270, 119
\bibitem{}Ramsay, G., Hakala, P., Wu, K., Cropper, M., Mason, K. O., 
C\'{o}rdova, F. A., Priedhorsky, W., MNRAS, 2005, 357, 49
\bibitem{}Ramsay, G., Cropper, M., Hakala, P., 2006, MNRAS, 367, L62
\bibitem{}Schwope, A. D., Hambaryan, V., Haberl, F., Motch, C., 2007,
Ap\&SS, 308, 619
\bibitem{}Steeghs, D., Marsh, T. R., Barros, S. C. C., Nelemans, G., 
Groot, P. J., Roelofs, G. H. A., Ramsay, G., Cropper, M., 2006, ApJ, 649, 382
\bibitem{}Wilms, J., Allen, A., McCray, R., 2000, ApJ, 542, 914
\bibitem{}Wu, K., Cropper, M., Ramsay, G., Sekiguchi, K., 2002, MNRAS, 331, 221
\bibitem{}Zane, S., Cropper, M., Turolla, R., Zampieri, L., Chieregato, M.,
Drake, J. J., Treves, A., 2005, ApJ, 627, 397

\end{thebibliography}
\end{document}